# Is Human Culture Locked by Evolution?

Hao Wang

*Ratidar.com*
*haow85@live.com*

**ABSTRACT**
Human culture has evolved for thousands of years and thrived in the era of Internet. Due to the availability of big data, we could do research on human culture by analyzing its representation such as user item rating values on websites like MovieLens and Douban. Industrial workers have applied recommender systems in big data to predict user behavior and promote web traffic. In this paper, we analyze the social impact of an algorithm named ZeroMat to show that human culture is locked into a state where individual's cultural taste is predictable at high precision without historic data. We also provide solutions to this problem and interpretation of current Chinese government's regulations and policies.

## 1. INTRODUCTION

Human culture represents the most valuable human imagination and emotions. It is the memory of the public psychology and aesthetics. Researchers have spent thousands of years to establish the value systems and methodologies of human culture including fine arts, music, literature, and so on. However, due to the geographical confinement and poor communication channels, research on human culture on large scale dataset was not available until recent decades when logs of behaviors of the public can be recorded by big data on the internet.

In the internet era, we are able to collect collective culture interest of the public by internet products such as Goodreads, Douban, MovieLens [1], etc. We can investigate into the human culture by reviewing the user ratings on movies in the MovieLens dataset, or user ratings on books at Goodreads. One important area of this type of research is recommender systems. Recommender systems use algorithms to compute potentially interesting items to users based on historic data of the users on the website. The precision of the recommender system has been improving for the past three decades. Most people believe we can predict what users want to purchase with a profitable accuracy based on recommender systems.

In 2021, an algorithm named ZeroMat [2] was invented to tackle the cold-start problem for recommender systems. The cold-start problem refers to the difficulties of recommendation when a new user joins the website or a new item is introduced. In this context, no historic data can be utilized to predict whether the new user would like a specific item, or whether a new item would be liked by existing users. Most researchers rely on side information to solve this problem, or using heuristics to recommend. ZeroMat was the first algorithm that uses no input data except the maximum user rating value as predefined by the product owners. When compared with classic matrix factorization approach and the random guess heuristics, ZeroMat achieves pretty competitve results on MovieLens 1 Million Dataset (Fig.1 and Fig. 2):

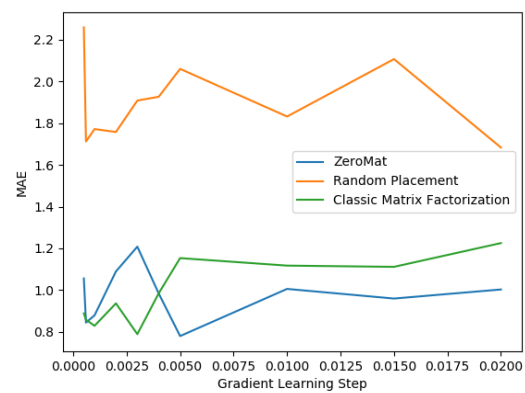

**Fig.1** Comparison in MAE

Since ZeroMat takes advantage only of parameter computations in its stochastic gradient descent process. Since MovieLens 1 Million Dataset represents a large sample of human movie interest, we can safely draw the

 

following conclusion: Human movie interest can be predicted at detailed levels with small errors without historic dataset. The evolution of human movie culture is locked into a history-irrelevant state if no government interference or regulation gets involved.

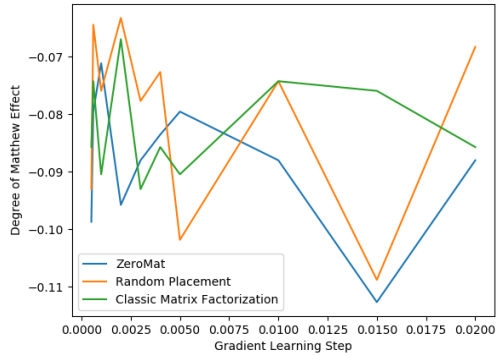

**Fig.2** Comparison in Degree of Matthew Effect

## 2. RELATED WORK

Zipf Distribution was a well-known probability distribution first discovered in Linguistics. It is a commonly encountered phenomenon in the field of social sciences ([3], [4]) and artificial intelligence ([5], [6]). However, its application in the field of recommender system is very recent.

Relatively unknown to the circles of social sciences and humanities, recommender system is an important technology and product in IT industry. Products such as "Guess what you like" on Taobao or news recommendation in Toutiao and video recommendation in TikTok all rely on recommender systems as the technical backbones of the products.

The discussion of our paper is based on social science analysis of a newly invented algorithm called ZeroMat. ZeroMat is an effective cold-start problem solver for recommender systems, and what is important, it is the first such algorithm in history that doesn't need any historical data. The success of ZeroMat is due to its master of Zipf Distribution, whose application in the recommender systems ([7], [8]) is only recent.

There is one milestone paper on Zipf Distribution. In 2021, Mazzarisi et.al. [9] invent an explainary theory of Zipf Distribution phenomenon: Zipf distribution is a consequence of maximization of efficiency in the system.

We explain the idea of lock-state of human culture in the context of big data of websites that collect millions of user data and culture item data. To be specific, we analyze ZeroMat's social implications on the MovieLens dataset. MovieLens dataset is one of the most popular open datasets used in the field of recommender systems. It contains large collections of users' ratings of movies on the website.

## 3. ZEROMAT AND LOCK-STATE

We discover that with no external intervention exerted, human culture will evolve to a lock-state, in which individual's or collective taste of cultural products can be calculated with high precision without using historical data. The reason of the lock-state of human movie culture is due to the underlying mechanism of Zipf distribution that dominates the cultural industry (or, user item rating distributions in technical details).

Zipf distribution, as well known in the field of Linguistics is the phenomenon observed in human language that the i-th most popular word has its frequency proportional to $\frac{1}{i}$. We discovered the lock-in state and its mechanism by analysis of ZeroMat algorithm. In ZeroMat, the researcher assume the user rating values follow Zipf distribution, namely most movies receive low scores while only a handful of movies receive high scores. Zipf distribution is a pretty close approximation to the real distribution and it exists not only in movie rating datasets, but also e-commerce datasets, music datasets, etc. This is a common-sense knowledge in the field of recommender systems.

ZeroMat builds its framework upon a recommendation algorithm named Matrix Factorization. In Matrix Factorization, the user item rating matrix is considered as factorization of the dot product of user feature vectors and item feature vectors. By computing these sets of vectors based on the incomplete historical data of user item rating values, we are able to construct unknown values in the full matrix.

ZeroMat specifies the distribution of user rating values in the following way :

$$\frac{R_{i,j}}{R_{max}} \sim \frac{U_i \bullet V_j}{\max(U_i \bullet V_j)} \quad (1)$$

, where R represents the user rating values, U means the user feature vector and V means the item feature vector. The idea of user and item feature vectors is borrowed from matrix factorization. Taking advantage of the probabilistic matrix factorization framework [10] , we acquire the following formulation :

$$P(U,V \mid R, \sigma_U, \sigma_V) = \prod_{i=1}^{N}\prod_{j=1}^{M}\left(U_i^T \bullet V_j\right) \times \prod_{i}^{N} e^{\frac{-U_i^T \bullet U_i}{2\sigma_U^2}} \times \prod_{i}^{M} e^{\frac{-V_i^T \bullet V_i}{2\sigma_V^2}} \quad (2)$$

The log of the probability is computed as below:

$$L = \sum_{i=1}^{N}\sum_{j=1}^{M}\ln(U_i^T \bullet V_j) - \frac{1}{2\sigma_U^2} \times \sum_{i=1}^{N} U_i^T \bullet U_i - \frac{1}{2\sigma_V^2} \times \sum_{i=1}^{M} V_i^T \bullet V_i \quad (3)$$

Optimization using stochastic gradient descent, we obtain the following parameter update rules (To simplify the model, we set standard deviations to be a constant) :

$$U_i = U_i + \eta \times \left( \frac{V_j}{U_i^T \bullet V_j} - 2 \times U_i \right) \quad (4)$$

$$V_j = V_j + \eta \times \left( \frac{U_i}{V_j^T \bullet U_i} - 2 \times V_j \right) \quad (5)$$

Obviously the update rules involve no historic data. To construct the unknown future user rating values in the cold-start case, we adopt the following formula:

$$R_{i,j} = R_{max} \times \frac{U_i \bullet V_j}{\max(U_i \bullet V_j)} \quad (6)$$

We compare the prediction of ZeroMat on unknown user item rating values with classic Matrix Factorization method with historic input data. Shockingly, ZeroMat outperforms classic Matrix Factorization in both accuracy (Fig. 1) and fairness (Fig. 2).

By reviewing the theory of ZeroMat, we proves that without external intervention, human movie culture will be locked up into a predictable state in which no historic data is relevant. The shocking conclusion is that since Zipf distribution exists nearly everywhere from human books, human movies, to human music, even human wealth distribution, our human society will be locked up into an extremely unfair state if freedom and time do their job.

Let's review something out of the cultural scope as well, we all know the degree distribution of social networks follow power law as well. No interference means human relations will end up in an extremely unfair state after evolution for some time.

Absolute freedom of human society means slavery in an extremely unfair state by evolution. We discuss the implications of ZeroMat and lock-state in the following section.

## 4. IMPLICATIONS

We observe the lock-state effect from the result of a cold-start AI algorithm, namely ZeroMat. ZeroMat proves that by mathematical calculation not relevant to historical data, one can predict the future trend of movie taste of the public at the individual level, achieving an accuracy level competitive with AI algorithm with full historical data.

As we discussed in the previous section, the reason of this precise prediction is due to the properties of the underlying probability distribution of the collective movie taste of the public. A natural implication is that if we disturb the underlying distribution of movie taste, or in a larger picture, human culture, we would be able to break the lock-state of human culture.

This idea is in fact already implemented by governments across the globe. Chinese government set a bar to the highest income and wages of movie stars, so the Zipf distribution related to movie stars was disturbed and it became less likely for the movie industry to fall into the lock-state.

If we take a look into an even broader spectrum of our human society, we will notice that Zipf distribution (or Power law distribution) is not only limited to cultural industry, but also city wealth distribution, human social network friendship, among many other scenarios.

One of the Chinese government's initiatives not too long ago was rapid urbanization, including the plan to create mega-metropolitan areas that grow faster than other places and entice more talents and investments, so a small cluster of mega-metropolitan areas could boost the overall economy of China.

The decision of the Chinese government is in fact correct because by [6], power law phenomenon is the consequence of maximization of efficiency. If we want more efficiency, the power law distribution needs to be more skewed. By increasing the skewness of city wealth distribution (i.e., making the wealth distribution biased even more towards richer cities) with mega-metropolis first policy, the Chinese government would increase the efficiency of the overall economy, which is the No.1 national goal for China in the past and envisable future.

Power law distribution is also ubiquitous in social networks. Scientists have discovered that social networks in Facebook, Twitter, Renren, and Douban follow power law distribution in degree distribution. This implies that the most sociable person has a disproportional larger number of friends than a less sociable person. If government does not intervene to reduce the level of the popularity of the most popular people in our society, the distribution of social tie resources would become dangerously biased towards a small handful of people. Several years ago, Chinese government regulated the circles of Public Intellectuals (公知/大 V in Chinese) to reduce the dissemination of fraud or dangerous information. The act is also justified by our analysis.

## 5. CONCLUSION

In this paper, we discuss a phenomenon that most people would find shocking - human culture will evolve into a lock-state after some time of evolution if no external intervention is exerted. The reason for the existence of the phenomenon is that Zipf distribution is the underlying mechanism of human culture taste distribution.

We suggest that governmental intervention and policies be implemented to eliminate the lock-state of human culture. We also give interpretation of the Chinese government's policy on regulation in the field of entertainment and public knowledge dissemination.

In future work, we would like to explore other possible explanation of the existence of lock-state of human culture. We would analyze ZeroMat in depth and in contrast to algorithms based on historical data set and provide a possible different interpretation of the validity and superiority of the algorithm. We hope in doing so we would gain more insight in social sciences.